\documentclass[showpacs,preprintnumbers,amsmath,amssymb]{revtex4}
\usepackage{graphicx}% Include figure files
\usepackage{dcolumn}% Align table columns on decimal point
\usepackage{bm}% bold math
\usepackage{natbib}
\usepackage{color}

\begin{document}

%\preprint{APS/123-QED}

\title{Hypocenter interval statistics between successive earthquakes \\in the two-dimensional Burridge-Knopoff model}% Force line breaks with \\

\author{Tomohiro Hasumi}
 \email{t-hasumi.1981@toki.waseda.jp}
%\author{Takuma Akimoto}
% \email{akimoto@aoni.waseda.jp}	
%\author{Yoji Aizawa}
% \email{aizawa@waseda.jp}
\affiliation{Department of Applied Physics, Advanced School of Science and Engineering, Waseda University, Tokyo 169-8555, Japan}

\date{\today}% It is always \today, today,

\begin{abstract}
We study statistical properties of spatial distances between successive earthquakes, the so-called hypocenter intervals, produced by a two-dimensional (2D) Burridge-Knopoff model involving stick-slip behavior. 
It is found that cumulative distributions of hypocenter intervals can be described by the $q$-exponential distributions with $q<1$, which is also observed in nature. 
The statistics depend on a friction and stiffness parameters characterizing the model and a threshold of magnitude. 
The conjecture which states that $q_t+q_r \sim 2$, where $q_t$ and $q_r$ are an entropy index of time intervals and spatial intervals, respectively, can be reproduced semi-quantitatively.  
It is concluded that we provide a new perspective on the Burridge-Knopoff model which addresses that the model can be recognized as a realistic one in view of the reproduction of the spatio-temporal interval statistics of earthquakes on the basis of nonextensive statistical mechanics.
\end{abstract}

%Use showkeys class option if keyword
\maketitle

\section{Introduction}
Earthquakes occur as a result of a fracture process and a frictional slip of a fault and are categorized into nonlinear complex phenomena in a non-equilibrium open system. 
In physics, the theory of complex systems and non-equilibrium systems has not been established, while in seismology the nature of a friction force acting on fault surfaces has not been elucidated. 
Thus, many fundamental problems remain unsolved. 
On the contrary, statistical features of earthquakes are well recognized as empirical laws, for example the Gutenberg-Richter (GR) law~\cite{Gutenberg:AG1956}, and the Omori law~\cite{Omori:JCSI1894}. 
As for spatio-temporal intervals between successive earthquakes, unified scaling laws were proposed~\cite{Corral:PRL2004, Davidsen:PRL2004}.

%Main suggested that the Earth is in a state of self-organized criticality (SOC)~\cite{Main:1997}, an idea originally~\cite{Bak:1988,Bak:1989}. In the concept of SOC, a system drives itself spontaneously into a stationary steady state, where spatio-temporal correlation functions exhibit a power law. Analogously, earthquakes show several types of power laws. The most established law is the Gutenberg-Richter (GR) law~\cite{Gutenberg:1956}, which stresses that the number of earthquakes $n$ with seismic magnitude grater than or equal to $m$ statistically has the form, $\log_{10} n = a - bm$, with positive constants $a$ and $b$, where $b$ is called the $b$-value, ranging from 0.80 to 1.1~\cite{Schorlemmer:2005}. The GR law can be transferred into a power-law for the probability density distribution of the earthquake size or the rupture area, $P(S) \propto S^{\delta}$. The exponent $\delta$ ranges between 0.80 and 1.05 in different faults~\cite{Pacheco:1992}. \par 
Inspired by the self-organized criticality (SOC)~\cite{Bak:PRA1988}, many earthquake models have been proposed and then compared the statistical features of events with those of earthquakes in nature~\cite{Rundle:AGU2000}. 
For instance the spring-block model proposed by Burridge and Knopoff~\cite{Burridge:BSSA1967} (hereafter referred to as the BK model) is well known as a simplified model of fault systems. 
Since Carlson and Langer presented a magnitude distribution which was similar to the GR law in one-dimensional (1D) BK model~\cite{Carlson:PRA1989}, many simulation studies based on this model have been carried out. 
Recent works of this model are focused on the spatio-temporal correlations~\cite{Mori:PRL2005, Mori:JGR2008}, interoccurrence time statistics, statistical properties of time intervals between successive earthquakes~\cite{Hasumi:PRE2007}, and a long range interaction~\cite{Xia:PRL2005}, and so forth. 
Although the BK model simplifies complex fault dynamics, the model is useful for discussion of statistical properties of earthquakes because the model reproduces the major statistical features of earthquakes, the GR law and interoccurrence time statistics~\cite{Hasumi:PRE2007}.\par
Nonextensive statistical mechanics proposed by Tsallis~\cite{Tsallis:JSP1988} and its applications have been paid much attention because this statistical mechanics is expected to provide a unified framework for understanding the statistical properties of spatio-temporal correlated systems and complex systems.
Nevertheless, earthquakes are phenomena exhibiting strong spatio-temporal correlations and great complexity. 
Hence the statistical properties of earthquakes based on this idea have been proposed, such as the modified GR law~\cite{Costa:PRL2004, Silva:PRE2006, Vilar:PA2007}, interoccurrence time statistics~\cite{Abe:PA2005, Darooneh:PA2008}, epicenter interval statistics~\cite{Abe:JGR2003}, and scale-free network topology~\cite{Abe:EPL2004}.
Especially, Abe and Suzuki found that the cumulative distribution of interoccurrence times and epicenter intervals can be described by the $q$-exponential distribution with $q>1$~\cite{Abe:PA2005} and with $q<1$~\cite{Abe:JGR2003}, respectively by analyzing Japan and Southern California earthquake data. 
Then they proposed a conjecture, which states that the sum of an entropy index obtained from the interoccurrence time distribution $q_t$ and that from the epicenter interval distribution $q_r$ is similar to two~\cite{Abe:JGR2003}. 
This relation is also observed by examining Iran earthquakes~\cite{Darooneh:PA2008}.
The present author reported that the cumulative distributions of interoccurrence times derived from the 2D BK model can be described by the $q$-exponential distribution with $q>1$, which reproduces the natural seismicity~\cite{Hasumi:PRE2007}. 
However, as far as we know, spatial interval statistics and the conjecture have not extracted yet.  
It is remarked that distributions of epicenters and network of epicenters in the Olami-Feder-Christensen model~\cite{Olami:PRL1992} were reported~\cite{Peixoto:PRE2004, Peixoto:PRE2006}.
They showed that these statistical features are similar to the observed behavior of real earthquakes~\cite{Abe:EPL2004}. \par
We try to work out how the hypocenter interval statistics are affected by the change of major physical quantities, for instance the friction force acting on the surface of faults, stiffness of a fault, and a threshold of magnitude. 
To achieve our aim statistical properties of the hypocenter intervals between successive earthquakes are investigated by analyzing synthetic data created by the 2D BK model.
Additionally, we examine the dependence of  the statistics on a friction and stiffness parameters, and the threshold of magnitude.  
Finally, the relation between $q_t$ and $q_r$ is discussed. 
We can show that the spatial interval statistics derived from the 2D BK model can reproduce semi-qualitatively that statistics in nature.

\section{Model}
\begin{figure*}[]
\begin{center}
\includegraphics[width=.4\linewidth]{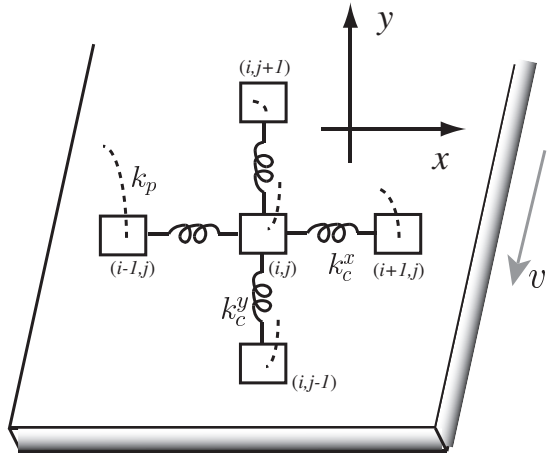}
\end{center}
\caption{Block and spring system for the 2D BK model. The model is composed of blocks of equal mass, $m$, of two different of coil springs, $k_c^x$ and $k_c^y$, and of the leaf spring, $k_p$. Each mass is subjected to the friction force, which depends only on the velocity of the block.}
\label{BK_dim2}
\end{figure*}
We display a diagram of the 2D BK model in Fig.~\ref{BK_dim2}. 
In this model, a fault is represented by the 2D network of the blocks interconnected by springs, $k_c^x, k_c^y$, and $k_p$. 
In this work, we assume that the slip direction of blocks is restricted to the $y$-direction. 
The equation of motion of the block at site $(i,j)$ is expressed by
\begin{eqnarray}
m\frac{d^2 y_{i,j}}{dt^2} &=& k_c^x(y_{i+1,j}+y_{i-1,j}-2y_{i,j})  + k_c^y(y_{i,j-1}+y_{i,j+1}-2y_{i,j})- k_p y_{i,j} -  F\left(v+\frac{dy_{i,j}}{dt}\right), 
\label{eqn:EQM}
\end{eqnarray}
where $m$ is the mass of the block, $y$ is displacement, and $F$ is a dynamical friction force as a function of a slipping velocity of a block. 
In order to rewrite Eq.~(\ref{eqn:EQM}) as a non-dimensional form, we define a dimensionless displacement $U$, a dimensionless dynamical friction force $\phi$, and dimensionless time $t'$, respectively, as 
\begin{eqnarray}
U_{i,j}= y_{i,j}/D_0 = y_{i,j}/(F_0/k_p),  F(\dot{y}_{i,j}) = F_0\phi(\dot{y}_{i,j}/v_1),  t' = \omega_pt = \sqrt{k_p/m}\; t, \nonumber
\end{eqnarray}
where $F_0$ is the maximum friction force and $v_1$ is the characteristic velocity at $F_0/2$. 
Using the parameters defined above and Eq.~(\ref{eqn:EQM}), we obtain
\begin{eqnarray}
\frac{d^2 U_{i,j}}{dt'^2} &=& l_x^2(U_{i+1,j}+U_{i-1,j}-2U_{i,j})  +  l_y^2(U_{i,j-1}+U_{i,j+1}-2U_{i,j}) - U_{i,j} -  \phi \left(2\alpha \left(\nu+\frac{dU_{i,j}}{d t'}\right)\right),
\label{eqn:noneqm}
\end{eqnarray}
where
\begin{eqnarray}
l_x = \sqrt{\frac{k_c^x}{k_p}}, \; l_y = \sqrt{\frac{k_c^y}{k_p}}, \; \nu = \frac{v}{\omega_pD_0} = \frac{v}{v^*}, \; 2\alpha = \frac{v^*}{v_1}. \nonumber 
\end{eqnarray}
$l_x$ and $l_y$ are the ratio of stiffness of the $x$ and $y$ directions, respectively. 
$\nu$ is a dimensionless loading velocity. 
$\alpha$ is the ratio of the maximum slipping velocity, $v^*$, to the characteristic velocity, $v_1$. 
We use the non-dimensional friction function $\phi(\dot{U})$, namely 
\begin{eqnarray}
\phi (\dot{U}) = \left\{
\begin{array}{ll}
(-\infty, 1] & \dot{U}=0,\\
{\displaystyle \frac{(1-\sigma)}{\{1+2\alpha[\dot{U}/(1-\sigma)]\}}} & \dot{U}>0,
\end{array}
\right.
\label{friction_function}
\end{eqnarray}
where $\alpha$ and $\sigma$ are control parameters. $\alpha$ represents the velocity-weakening tendency and indicates how quickly the dynamical friction force decreases with increasing velocity. 
Simultaneously, $2\alpha$ is the differential coefficient at $\dot{U}=0$. 
In the case for $\alpha=0$, the dynamical friction force is constant, $\phi(0)=1-\sigma$, and for large $\alpha$, the force decreases rapidly to 0. 
According to the relation between the friction force and the slip velocity based on the rock fracture experiment~\cite{Scholz:Book2002}, $\alpha$ is on order of unity. 
$\sigma$ is the difference between the maximum friction force (=1) and the initial stage of the dynamical friction force $(=\phi(0))$. 
In order to prevent a back slip, which means that blocks slip in the $-y$ direction, $\phi$ ranges from $-\infty$ to 1 at $\dot{U}=0$. \par

Previous works reported that the parameter regime, where the 2D BK model can be recognized as a realistic earthquake model, is limited and is estimated to be $l_x^2=1, l_y^2=3$, and $\alpha \approx3.5$~\cite{Hasumi:PRE2007, Kumagai:GRL1999}.
In this case, the model can extract the GR law with $b=1$~\cite{Hasumi:PRE2007, Kumagai:GRL1999}, a statistical property of stress drop~\cite{Kumagai:GRL1999}, the constant stress drop, the Zipf-Mandelbrot type power law for interoccurrence time statistics~\cite{Hasumi:PRE2007}, and the ratio of the seismic wave velocity~\cite{Hasumi:PRE2007}. \par

In this work, the system size is taken to be $N_x=100$ and $N_y=25$. 
As we described before, the 2D BK model is characterized by five parameters, $l_x^2, l_y^2, \alpha, \sigma$, and $\nu$. 
Here the two model parameters, $\sigma$ and $\nu$, are fixed to be 0.01, whereas $l_x^2, l_y^2$, and $\alpha$ are varied around the optimal parameter regime. 
It is confirmed that hypocenter statistics do not change quantitatively by varying $\sigma$ and $\nu$. 
We use the fourth-order Runge-Kutta algorithm with time step $\Delta t = 0.001$ to solve Eqs.~(\ref{eqn:noneqm}) and (\ref{friction_function}) under a free boundary condition. 
The initial block displacements have small irregularities. 
Earthquake-like events are used after a certain period of time when the initial random configurations do not influence statistical properties. A seismic magnitude $m$ in this model is defined as $m = \log_{10} \left( \sum _{i,j}^n  \delta u_{i,j} \right)/1.5$, where $\delta u_{i,j}$ stands for the total displacement at site $(i,j)$ during an event and $n$ is the number of slipping blocks.

\section{Results and Discussion}
We study hypocenter interval statistics between successive earthquakes. 
In this model, a position where a block slips for the first time during an event is considered as a hypocenter. 
The $n$th hypocenter distance is defined as $r_{n} = |\vec{{r}}_{n+1} - \vec{{r}}_{n}|$, where $\vec{{r}}_{n}$ and $\vec{{r}}_{n+1}$ are the position vector of the $n$th and $n+1$th earthquake's hypocenter, respectively. 
We introduce the scaled distance, $r'=r/\bar{r}$, where $\bar{r}$ is a normalized parameter and is set at 20.0, arbitrary. 
Actually, the statistics we will present later do not change statistically by varying $\bar{r}$. \par
We focus our attention on the applicability of the $q$-exponential distribution for the cumulative distribution of hypocenter intervals $P(>r')$. 
According to the paper~\cite{Abe:JGR2003}, $P(>r')$ can be described by the $q$-exponential distribution, $e_q(x)$, namely 
\begin{equation}
P(>r') =  e_q(-r'/r_0) = [1+(1-q)(-r'/r_0))^{1/(1-q)}]_+,  
\label{eqn:power_d}
\end{equation}
where $q$ and $r_0$ are respectively called the entropy index and the length-scale parameter, and $[a]_+ \equiv$ max $[0,a]$. 
It is known that the $q$-exponential distribution converges to the exponential distribution as $q \rightarrow 1$, and that for $q>1$ the $q$-exponential distribution corresponds to the power law~\cite{Tsallis:JSP1988}. 
The inverse function of $e_q(x)$ is called the $q$-logarithmic function, $\ln_q(x)$, given by 
\begin{equation}
\ln_q (x) = \frac{1}{1-q}(x^{1-q} -1).
\end{equation}
Then, the hypocenter interval statistics are studied produced by the 2D BK model for different $l_x^2, l_y^2, \alpha$, and the threshold of magnitude $m_c$. 
In this study, we select the $q$-exponential distribution with $q<1$ for the ideal curve of $P(>r')$, because the probability density function of the hypocenter intervals in this model does not obey the power law. 

\subsection{Friction parameter $\alpha$ dependence} 
\begin{figure*}[]
\begin{center}
\includegraphics[width=.95\linewidth]{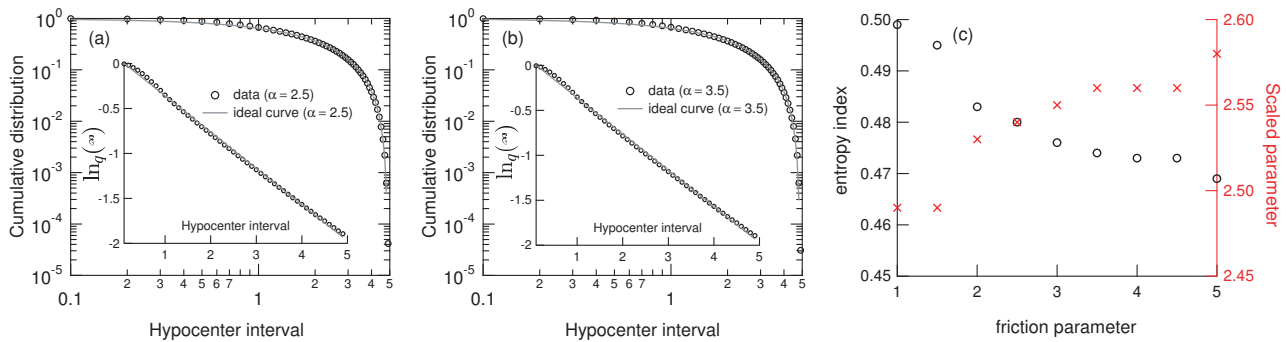}
\end{center}
\caption{Cumulative distributions of hypocenter intervals between successive events for different $\alpha$, whereas $l_x^2=1$ and $l_y^2=3$. The parameter $\alpha$ is $\alpha=2.5$ (a) and 3.5 (b). The makers and the solid lines correspond to the numerical results of $P(>r')$ and the ideal curve of $P(>r')$ defined in Eq.~(\ref{eqn:power_d}). We demonstrate that the inset figures represent the semi-$q$-log plot of the data. The fitting parameters $q_r~(\circ)$ and $r_0~(\times)$ as a function of $\alpha$ are shown in (c).}
\label{friction}
\end{figure*}

First, we examine the dependence of the hypocenter interval statistics on the friction parameter $\alpha$. 
For that purpose the stiffness parameters, $l_x^2$ and $l_y^2$ are fixed ($l_x^2=1$ and $l_y^2=3$) and $\alpha$ is changed from 1.0 to 5.0. 
As shown in fig.~\ref{friction},(a) and (b), the cumulative distributions of hypocenter intervals are well fitted by the $q$-exponential distribution with $q<1$. 
For $\alpha=2.5$ and 3.5, the fitting parameters and the $r^2$-value in other words the correlation coefficient, denoted by, $R^2_r$ are estimated to be $q_r=0.479, r_0=2.54$ and $R^2_r=0.9988$ for (a) and $q_r=0.474, r_0=2.56$, and $R^2_r=0.9986$ for (b) by means of the least-squire root test. 
$q_r$ decreases as $\alpha$ increases, while $r_0$ increases as $\alpha$ increases (see fig.~\ref{friction} (c)). 
$R^2_r$ ranges from 0.9985 ($\alpha=4.5$) to 0.9991 ($\alpha=1.0$). 
It is found that the hypocenter interval statistics depend on $\alpha$.

\subsection{Stiffness parameters $l_x^2$ and $l_y^2$ dependence}

In our second performance of our simulation, the stiffness parameter dependence of hypocenter intervals is studied. 
In this time, $\alpha$ is fixed at 3.5, whereas $l_x^2$ and $l_y^2$ are systematically changed. 
Our results of fitting parameters and $R^2_r$ are listed in Table.~\ref{table1}. 
As can be seen from this table, for large $l_x^2$ and $l_y^2$, $q_r$ tends to be large, while $r_0$ comes to be small, indicating the fact that $P(>r')$ follows the $q$-exponential distribution and depends on the stiffness of the system. 
Thus, we can conclude that the cumulative distributions of hypocenter intervals can be expressed by the $q$-exponential distribution with $q<1$ by varying $l_x^2, l_y^2$ and $\alpha$. %This indicates the view that the model is able to reproduce semi-qualitatively the observed behavior of real earthquakes.  
\begin{table}
\caption{\label{table1}The results of fitting parameters of $P(>r')$ for different $l_x^2$ and $l_y^2$ with fixing $\alpha=3.5$.}
\begin{center}
\begin{tabular}{c|c|c|c|c}
\hspace{5mm} $l_x^2$\hspace{5mm} &\hspace{5mm} $l_y^2$\hspace{5mm} &
 \hspace{5mm} $q_r$\hspace{5mm} & \hspace{5mm}$r_0$ \hspace{5mm} &\hspace{5mm} $R^2_r$\hspace{5mm} \\
\hline
\hline
 0.5 & 2.5 & 0.472 & 2.56 & 0.9985  \\
\hline
 0.5 & 8 & 0.481 & 2.53 & 0.9990 \\
\hline
 1 & 1 & 0.473 & 2.56 & 0.9904 \\
\hline
 1 & 5 & 0.474 & 2.56 & 0.9987 \\
\hline
 2 & 3 & 0.468 & 2.59 & 0.9987 \\
\hline
 3 & 3 & 0.469 & 2.59 & 0.9989 \\
\hline
 3 & 6 & 0.480 & 2.55 & 0.9990 \\
\hline
 4 & 8 & 0.490 & 2.51 & 0.9989 \\
\hline
 5 & 8 & 0.494 & 2.49 & 0.9989 \\
\hline
 10 & 30 & 0.508 & 2.42 & 0.9988 \\
\hline
\end{tabular}
\end{center}
\end{table}

\subsection{Threshold of magnitude $m_c$ dependence}
\begin{figure*}[]
\begin{center}
\includegraphics[width=.4\linewidth]{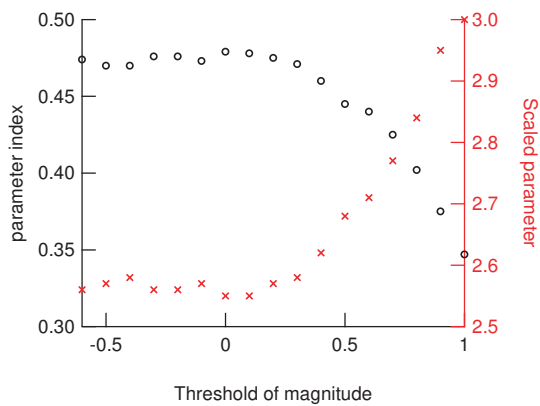}
\end{center}
\caption{Dependence of $q_r$ and $r_0$ on the value of threshold of magnitude $m_c$. The data is employed  in the case of  $l_x^2=1, l_y^2=3$ and $\alpha=3.5$.}
\label{magnitude}
\end{figure*}

Thirdly, the hypocenter statistics above a certain magnitude $m_c$ are examined. 
We use the data with $l_x^2=1, l_y^2=3$ and $\alpha=3.5$ because in this case the model yields the realistic magnitude distributions and interoccurrence time statistics~\cite{Hasumi:PRE2007}. 
For the parameter settings given above the magnitude $m$ ranges from $-0.65$ to 1.5. 
To keep the statistical quantities accurate, we determine the upper limit of $m_c$ to be 1.0. 
Then hypocenter interval statistics are studied by varying $m_c$ from $-0.6$ to 1.0. 
In fig.~\ref{magnitude} we display a change of fitting parameters $q_r$ and $r_0$ as a function of $m_c$. 
At first, $q_r$ and $r_0$ do not change in the magnitude region, $-0.60 < m_c \lessapprox 0.1$. 
However, for $0.1 \lessapprox m_c \le 1.0$ as $m_c$ increases, $q_r$ decreases and $r_0$ increases. 
In this case the value of $R^2_r$ is between 0.9924 (for $m_c=-0.60$) and 0.9986 (for $m_c=1.0$) so that the $q$-exponential distribution is suitable to describe $P(>r')$ in the magnitude domain, $-0.6 \le m_c \le 1.0$. 

\subsection{System size dependence}
\begin{figure*}[]
\begin{center}
\includegraphics[width=.4\linewidth]{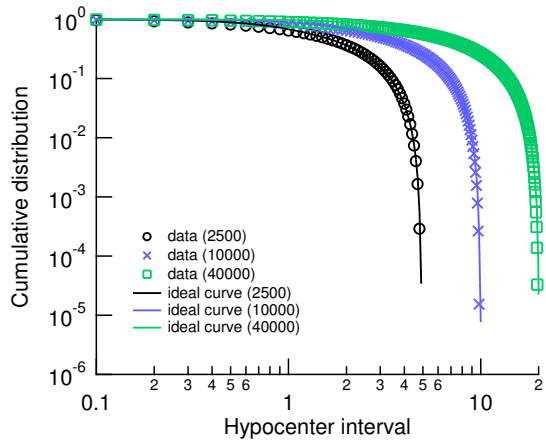}
\end{center}
\caption{The cumulative distribution of hypocenter intervals between successive events for different $N$, whereas $l_x^2=1, l_y^2=3$, and $\alpha=3.5$. The markers and the solid line correspond to the numerical data and ideal curve of $P(>r')$.}
\label{size}
\end{figure*}

Now, we focus on the system size dependence of the hypocenter interval statistics. 
We change the number of blocks $N$ from $2500~(25 \times 100)$ to $40000~(100 \times 400)$, with fixing $l_x^2=1, l_y^2=3$, and $\alpha=3.5$; $N$ is taken to be $N=2500~(25\times 100)$, $N=10000~(50\times 200)$, and $N=40000~(100 \times 400)$ (see Fig.~\ref{magnitude}). 
As shown in fig.~\ref{size}, the cumulative distribution of hypocenter intervals $P(>r')$ is well fitted by the $q$-exponential distribution with $q<1$ for all the cases. 
The fitting parameters and $R^2_r$ are estimated to be $q_r=0.474, r_0=2.56$, and $R^2_r=0.9986$ for $(N=2500)$, $q_r=0.474, r_0=5.17$, and $R^2_r=0.9985$ for $(N=10000)$, and $q_r=0.477, r_0=10.35$, and $R^2_r=0.9984$ for $(N=40000)$. 
$q_r$ do not depend on $N$, whereas $r_0$ increase linearly with $N$; $r_0$ for $N=10000$ is about twice as large as that for $N=2500$, and $r_0$ for $N=40000$ is about four times as large as that for $N=2500$. 
Thus it is concluded that the hypocenter interval statistics hold up when the system size increases in the range $2500 \lessapprox N \lessapprox 40~000$. 

\subsection{Conjecture}
Finally, we check the conjecture demonstrated in Ref.~\cite{Abe:JGR2003}, which states that $q_t+q_r \sim 2$. 
The results of spatio-temporal interval statistics by analyzing the synthetic data for different  $l_x^2, l_y^2$, and $\alpha$ and earthquake data of Southern California~\cite{Abe:PA2005, Abe:JGR2003}, Japan~\cite{Abe:PA2005, Abe:JGR2003}, and Iran~\cite{Darooneh:PA2008} are listed in Table~\ref{table2}. 
$R^2_t$ in this table means the $r^2$-value of interoccurrence time statistics. 
Note that we only focus on the case of control parameters where the $q$-exponential distribution is more suitable distribution for ideal curve of $P(>\tau)$ than the exponential distribution $(q=1)$.
In addition, for $l_x^2=1, l_y^2=3$, and $\alpha=3.5$, the value of  $q_t$ and $R^2_t$ are quoted by our previous paper~\cite{Hasumi:PRE2007}. 
As shown in Table~\ref{table2}, it can be deduced that the sum of $q_t$ and $q_r$ in this model shows $q_t+q_r \sim 1.5$, which is similar to the observed value. 

\begin{table}
\caption{\label{table2}Summary of the results of fitting parameters of spatio-temporal intervals between successive events based on the 2D BK model and on earthquake data.}
\begin{center}
\begin{tabular}{ccccccc|c}
\hspace{5mm}$l_x^2$\hspace{5mm} &\hspace{5mm}$l_y^2$\hspace{5mm} &\hspace{5mm}$\alpha$\hspace{5mm} &
\hspace{5mm} $q_r$\hspace{5mm} &\hspace{5mm} $R^2_r$\hspace{5mm} &\hspace{5mm} $q_t$\hspace{5mm} &
\hspace{5mm} $R^2_t$ \hspace{5mm} &\hspace{5mm}$q_r+q_t$ \hspace{5mm} \\
\hline
\hline
 0.5 & 0.5 & 2.0 & 0.472 & 0.9982 & 1.01 & 0.9953 & 1.482 \\
\hline
 0.5 & 1 & 2.0 & 0.472 & 0.9904 & 1.10 & 0.9984 & 1.572 \\
\hline
 1 & 3 & 3.5 & 0.474 & 0.9986 & 1.08 & 0.9890 & 1.554 \\
\hline
 1 & 3 & 4.5 & 0.473 & 0.9985 & 1.04 & 0.9950 & 1.513 \\
\hline
 1 & 2 & 4.0 & 0.468 & 0.9982 & 1.02 & 0.9960 & 1.488 \\
\hline
 1.5 & 2 & 4.0 & 0.471 & 0.9983 & 1.03 & 0.9890 & 1.501 \\
\hline
 3 & 3 & 10 & 0.469 & 0.9982 & 1.10 & 0.9832 & 1.569 \\
\hline
 4 & 5 & 10 & 0.468 & 0.9982 & 1.07 & 0.9948 & 1.538 \\
\hline
 \multicolumn{3}{c}{Southern California}~\cite{Abe:PA2005, Abe:JGR2003} & 0.773 & 0.9993 & 1.13 & 0.98828 & 1.903 \\
\hline
 \multicolumn{3}{c}{Japan}~\cite{Abe:PA2005, Abe:JGR2003} & 0.747 & 0.9990 & 1.05 & 0.99007 & 1.797 \\
\hline
 \multicolumn{3}{c}{Iran}~\cite{Darooneh:PA2008} & 0.817 & 0.9970 & 1.308 & 0.998 & 2.125 \\
\hline
\end{tabular}
\end{center}
\end{table}

\section{Summary}
We have found a new insight into the 2D BK model in connection with the hypocenter interval statistics between successive earthquakes on the basis of the nonextensive statistical mechanics. 
It is shown that the cumulative distribution of hypocenter intervals is in agreement with the $q$-exponential distribution with $q<1$, which reproduces the observed behavior of real earthquakes.
The statistics depend on the dynamical parameters, stiffness of the system, $l_x^2$ and $l_y^2$ and frictional features of a fault, $\alpha$, and the threshold of magnitude $m_c$. 
In detail, the distribution function do not change, while the fitting parameters $q_r$ and $r_0$ depend on $l_x^2, l_y^2, \alpha$, and $m_c$. 
Additionally, we have checked the conjecture, which is a relation between $q_t$ and $q_r$ and revealed that $q_t + q_r \sim 1.5$. 
Although the physical interpretation of this relation remains open, our findings can lead to the conclusion that the model is able to extract the spatial interval statistics in nature semi-qualitatively. \par
It is known that the BK model is highly simplified, so that many features of the fault system have been neglected. 
However, the model can be recognized as a realistic model in view of the reproduction of statistical features of spatio-temporal intervals between successive earthquakes based on nonextensive statistical mechanics. 
In a future publication, we will focus on the network of epicenters in this model.  
Finally, we hope that this work is a first step toward understanding fully the origin of statistical properties of earthquakes and of other physical systems exhibiting the stick-slip motion. \par

%The author would like to thank Professor Y. Aizawa and Dr. T. Akimoto for discussions. 

\begin{acknowledgments}
The author would like to thank Prof. Y. Aizawa and Dr. T. Akimoto for encouragement and Dr. M. Kamogawa, Prof. H. Kawamura, and Dr. T. Hatano for useful comments and helpful discussions. 
This work is partly supported by the Sasagawa Scientific Research Grant from The Japan Science Society. 
The author is grateful for research support from the Japan Society for the Promotion of Science (JSPS) and the Earthquake Research Institute cooperative research program at the University of Tokyo. 
Thanks are also extended to one anonymous reviewer for improving the manuscript. 
\end{acknowledgments}

\end{document}